\documentstyle[prc,preprint,tighten,aps]{revtex}

\title{\bf A Neutral Atom and a Charged Wire: From Elastic Scattering to
Absorption }

\author{M.\ Bawin\thanks{Electronic-mail address: michel.bawin@ulg.ac.be}}
\address{Universit\'{e} de Li\`{e}ge, Institut de Physique B5, Sart 
Tilman, 4000 Li\`{e}ge 1, Belgium  }

\author{S.\ A.\ Coon\thanks{Electronic-mail address: coon@nmsu.edu}}
\address{Physics Department, New Mexico State University,
         Las Cruces, NM  88003, USA}

\begin{document}

\maketitle

\begin{abstract}    
We solve the problem of a neutral atom interacting with an   charged
wire, giving rise to an attractive   $1/r^2$ potential in two
dimensions.  We show how a suitable average over all possible
self-adjoint extensions of the  radial Schroedinger Hamiltonian
eventually leads to the classical formula for absorption of the atom, a
formula shown to be in agreement with a recent experiment.

\end{abstract}


\vskip20pt

The scattering of cold neutral atoms by a thin charged wire has, for
the first time, allowed the {\em experimental} study of a pure
attractive  $1/r^2$ potential in two dimensions\cite{PRL_98}.  The
absorption aspects of this experiment were  successfully described by a
solution of the classical equations of motion. It is the purpose of
this paper to show how this absorption formula can be derived from the
{\em elastic} scattering solutions of the Schroedinger equation for
this attractive singular potential\cite{singular} instead of introducing {\em ad hoc}  complex phase
shifts\cite{DS_97,ASV_99}   To do this, we employ a  suitable
average of the  $S$-matrix arising from the mathematically well defined
solutions of the Schroedinger equation with a self-adjoint Hamiltonian
and compute the corresponding absorption cross-section in the classical
limit.  Our absorption cross-section is then identical to the classical
absorption cross-section which describes the data of Ref.
\cite{PRL_98}.

To begin, we note that the electrical field of a wire with line charge
per unit length $\lambda$ induces a dipole moment $\vec{d} = \alpha
\vec{E}$ in a
neutral atom of polarizability  $\alpha$  which is then attracted towards
the wire.  The interaction potential in cylindrical coordinates (and
Gaussian units so that the fine structure constant $e^2/\hbar c \approx 1/137$), 

 \begin{equation}
V{\rm_{pol}}(r) = -\, \frac{1}{2} \vec{d}\cdot\vec{E} = 
-\, \frac{1}{2} \alpha E^2(r) = -\,\frac{2 \alpha \lambda^2}{r^2}\, ,
\label{eq:vpol}
\end{equation}
is always attractive. 
The radial Schroedinger Hamiltonian for the atom with mass $M$ is
 
\begin{equation}
H= -\left\{{1\over 2M}\left[{\partial^2\over\partial r^2}
+{1\over r}{\partial\over\partial r}
-{m^2\over r^2}\right]
+{2\alpha \lambda^2\over  r^2}\right\}\, ,  \label{eq:ham}
\end{equation}
and the radial Schroedinger equation then becomes
 
\begin{equation}\label{eq:Se}
\left\{{1\over 2M}\left[{\partial^2\over\partial r^2}
+{1\over r}{\partial\over\partial r}
-{m^2\over r^2}\right]
+{2\alpha \lambda^2\over  r^2} + {\cal E}\right\} \psi(r) = 0\,,
\label{eq:sch}
\end{equation}
where $m$, the orbital angular momentum quantum number in the
$z$-direction (direction of the wire), takes the values $0,\pm 1,\pm 2,
\ldots$ and ${\cal E}$ is the energy of the atom ($\hbar=c=1$).  We
require $\nu^2 \equiv 4 \alpha \lambda^2 M -m^2 $ to be greater than
zero, so that the singular potential remains attractive.  We
follow Meetz\cite{Meetz} and define a two term regular\cite{regular}
solution $\phi$ for this singular potential
 \begin{equation}
   \phi(kr) = C [e^{i\gamma} J_{i\nu}(kr) - e^{-i\gamma}J_{-i\nu}(kr)]
   \,,  \label{eq:phi}
\end{equation}
with $k^2/2M= {\cal E}$ , $\nu^2 = 4 \alpha \lambda^2 M -m^2$, and
$\gamma$ is an arbitrary phase which characterizes the self-adjoint
extension\cite{Meetz,Case,Faris,ASV_00} of the radial Schroedinger
Hamiltonian (\ref{eq:ham}).
 
The partial wave $S$-matrix $S_m$ is given by \cite{Newton}
\begin{equation}
	S_m = \frac{\cal {L}_{-}}{\cal {L}_{+}}\,,  \label{eq:Smatrix}
\end{equation}
 where the Jost function $\cal {L}_{+}$ and $\cal {L}_{-}$ in
 (\ref{eq:Smatrix}) 
 are determined
   by
 the asymptotic behavior $r \rightarrow \infty$ of $\phi(kr)$. This in
 turn is easily found from\cite{Abramovitz}:

\begin{equation}
J_{\nu}(z) \sim \sqrt{\frac{2}{\pi z}} \cos\left( z -\frac{\nu \pi}{2} -
		\frac{\pi}{4}\right) \,. \label{eq:J_nu}
\end{equation}
Comparing (\ref{eq:phi}) and (\ref{eq:J_nu}), we now evaluate 
 the Jost function $\cal {L}_{+}$ and $\cal {L}_{-}$ 
 
\begin{equation}
      \phi(kr) \simeq
       \frac{C}{\sqrt{2kr}} ({\cal L}_{-}e^{ikr} - 
       {\cal L}_{+}e^{-ikr})\,,   \label{eq:Jost}
\end{equation}
and it is not necessary to determine the value of  
$C$ in (\ref{eq:phi}) or (\ref{eq:Jost})
as it does not appear in the partial wave $S$-matrix:
\begin{equation}
S_m(\gamma) = \frac{\cal {L}_{-}}{\cal {L}_{+}} = 
\frac{e^{i\gamma} e^{\frac{\nu \pi}{2}} - e^{-i\gamma} e^{-\frac{\nu
\pi}{2}} }  
     {e^{i\gamma} e^{-\frac{\nu \pi}{2}}-e^{-i\gamma} e^{\frac{\nu
     \pi}{2}}}\,\,.
\end{equation}

Now we utilize a method suggested by Radin\cite{Radin} to clarify the
relation between the
family of solutions in (\ref{eq:phi}),
 each solution characterized by $\gamma$\cite{Case}, 
 with the alternative unique solution,
displaying absorption, found by Nelson\cite{Nelson}, for the same
attractive  $1/r^2$ potential.  We, however, apply the method
to $S$-matrix elements rather than Radin's Green functions and we
work in two dimensions.
Thus, we average $S_m(\gamma)$ over {\em all} $\gamma$'s corresponding to 
{\em all} self-adjoint extensions:
\begin{equation}
 \langle S_m \rangle = \frac{1}{2\pi} \int_{0}^{2\pi} 
 S_m(\gamma) \,d\gamma \,,   \label{eq:ave}
\end{equation}
and now proceed to show that $\langle S_m \rangle $ displays absorption.
Setting $z=e^{i\gamma}$, the average in (\ref{eq:ave}) becomes  

\begin{equation}
	2\pi i \langle S_m \rangle =  e^{\frac{\nu \pi}{2}} 
	\oint_{|z|=1}  \frac  {dz} {e^{\frac{-\nu \pi}{2}} z 
	- e^{\frac{\nu \pi}{2}}/z}
		-e^{-\frac{\nu \pi}{2}}\oint_{|z|=1}
	\frac  {dz/z} {e^{\frac{-\nu \pi}{2}} z^2 
	- e^{\frac{\nu \pi}{2}}} \,, \label{eq:contour}
\end{equation}
and is readily evaluated as:
\begin{equation}
\langle S_m \rangle = e^{-\nu \pi} \,,  \label{sm}
\end{equation}
because the first contour integral of (\ref{eq:contour}) has 
no poles inside the contour.

The absorption cross-section is given in terms of the partial wave
expansion as\cite{DS_97,Gottfried}
\begin{equation}
     \sigma_{abs} = \frac{1}{ k}\sum_{-\mu}^{+\mu}  (1 -
     |S_m|^2)\,.
				\label{eq:absorb}
\end{equation}
Because $\nu^2 = 4 \alpha \lambda^2 M -m^2 > 0$, the limits of the sum
 on $m$ are given by the integer part of $2\lambda
\sqrt{\alpha M}$ :  $\mu = [2\lambda \sqrt{\alpha M}]$.

In order to  take the quantum mechanical expression (\ref{eq:absorb})
 to the classical limit, 
we follow a treatment given by 
Alliluev\cite{Alliluev} in three dimensions.  The  two dimensional
absorption cross-section
(\ref{eq:absorb}) goes to
\begin{equation}
\sigma_{abs} = \frac{2}{k} \int_{0}^{+\mu}[1 - 
e^{-2 \pi \sqrt{\mu^2 - m^2}}] dm \,,
	     \label{eq:absint}
\end{equation}
where the classical limit is given by $\mu =2\lambda\sqrt{\alpha M}\gg
1$, see Kayser\cite{Kayser}.  The second term in the integral of 
(\ref{eq:absint}) is,  totally negligible with respect to the first
term (one can show the second integral to be ${\cal O}
(1/\sqrt{\mu})$).   We finally get, from the classical limit of  our
pure quantum mechanical treatment of this singular potential
\begin{equation}
\sigma_{abs} = \frac{4\lambda\sqrt{\alpha M}}{k}  =
2\sqrt{\frac{4\alpha}{M} \frac{\lambda^2}{v^2} } \label{eq:result}
  \,,
\end{equation}
where $k = Mv$ and the final form of our result (\ref{eq:result}) is seen
to be identical to equation (3) (in the limit of zero radius of the
wire) of Ref. \cite {PRL_98}, obtained by a classical argument.

Now that we have shown how  the classical result for  absorption by a
zero-radius charged wire follows from quantum mechanics, it is
straightforward to include classical absorption of an atom with energy
${\cal E}$ by the finite radius 
$R_{\rm w}$ of the wire by introducing a phenomenological potential of
the form $-\beta/r^2$.  Classically,  those atoms whose impact
parameter $\rho$ does not exceed $\rho_{\rm max}= \sqrt{\beta/{\cal E}}$ will
be absorbed (``fall to the center")\cite{LL}. In two dimensions
$\sigma_{abs} = 2\rho_{\rm max}$. For the uncharged wire of radius
$R_{\rm w}$ one gets $\sigma_{abs} = 2R_{\rm w}$ for a phenomenological
strength $\beta=R^2_{\rm w}{\cal E}$.  With  $\beta = R^2_{\rm w}{\cal
E} + 2\alpha
\lambda^2$, it is clear that classically one has
\begin{equation}
\sigma_{abs} = 2\sqrt{R_{\rm w}^2
+\frac{4\alpha}{M} \frac{\lambda^2}{v^2} }  \,. \label{eq:adhoc}
\end{equation}
in agreement with (\ref{eq:adhoc}) in Ref. \cite{PRL_98}.   Note that
the phenomenological classical potential we adopt for the uncharged
wire with a finite radius is also singular, a choice ultimately
justified by the successful description of the data by (\ref{eq:adhoc})
in Ref. \cite{PRL_98}.

We have emphasized the rigorous mathematics of this problem to
illustrate the point that there is no difficulty, neither quantum
mechanically nor mathematically, with formulating the problem of the
interaction of a polarizable atom with a zero-radius line charge,
contrary to statements in the physics literature \cite{Hagen}. 
Attempts have been made to give  the continuous parameter $\gamma$ of
the self-adjoint extensions of the   radial Schroedinger Hamiltonian
with an attractive $1/r^2$ potential a physical
interpretation\cite{ASV_00,PP}. One such interpretation is briefly
discussed  in the appendix in conjunction with a different method of
solving  eq. (\ref{eq:sch}). However, given the  difficulty in trying
to select a unique self-adjoint extension  based upon physical
arguments\cite{ASV_00}, we instead have shown that, by considering {\em
all} self-adjoint extensions by averaging over $\gamma$, one can go
from a well formulated quantum mechanical elastic scattering solution
to the classical formula for absorption, a formula which fits the
experimental data.

\section*{Acknowledgments}
The work of M.B. was supported by the National Fund for Scientific
Research, Belgium and that of S.A.C. by NSF grant PHY-9722122.  

\appendix
\section*{} 

For completeness sake, we briefly discuss how the quantum mechanical
expression (\ref{sm}) can be derived from the approach of the authors
of Ref.\cite{ASV_00}. These authors write the physical 
solution \cite{regular} $R_m(kr)$ to equation (\ref{eq:sch}) as 
\begin{equation}
 R_m(kr) = a_m J_{-\nu}(kr) + b_m J_{\nu}(kr),    
 \label{eq:schr}
  \end{equation} 
where $a_m$ and $b_m$ are determined by requiring the radial
wavefunctions to form an orthogonal set.They then find $R_m(kr)$ to be
of the form
\begin{equation}
\label{rs3'} 
R_m(kr) = c_m \left[e^{i[\theta_m+2\mu\ln
(k/M)]}J_{-i\nu}(kr) +  J_{i\nu}(kr)\right]\,,
 \end{equation}
where $c_m$ is a normalization factor and $\theta_m$ a phase
characterizing the self-adjoint extension. From equations
(\ref{rs3'}) and (\ref{eq:J_nu}) one now finds the partial wave
$\tilde{S}_m$ to be given by
 \begin{equation} 
 \tilde{S}_m =
\frac{e^{i(\theta_m+2\nu log\frac{k}{M})}e^{\frac{-\nu \pi }{2}} +
e^{\frac{\nu \pi }{2}}}{e^{i(\theta_m+2\nu log\frac{k}{M})}e^{\frac{\nu
\pi }{2}}
 + e^{\frac{-\nu \pi }{2}}}\,\,, \label{smc}
\end{equation}
since the connection between $\tilde{S}_m$ and the
asymptotic form of $R_m(kr)$ is  \cite{ASV_00}
\begin{equation} 
\label{asymR} 
R_m(\rho) \rightarrow \sqrt{1\over
2\pi p\rho} \left[e^{-i(p\rho-\pi m-{\pi\over 4})} +\tilde{S}_m
e^{i(p\rho-{\pi\over 4})}\right] \,.
 \end{equation} 
From (\ref{smc}), one then finds the poles of $\tilde{S}_m$ (bound
states) to be of the form $k=i\kappa $, with
 \begin{equation}
\label{kap} 
\kappa = e^{\frac{\theta'_m-2\pi n}{2\nu}}\,\,, \quad
n=0, \pm 1, \pm 2, ...\,\,  ,
 \end{equation} 
and
 \begin{equation}
\label{thet}
 \theta'_m= \pi - \theta_m \,\,.
  \end{equation}
Equation (\ref{kap}) shows that the domain of  $\theta'_m$ is
\begin{equation} 
0\leq\theta'_m<2\pi\,\,, 
\end{equation}
in order to include all possible  values of $\kappa$ in Radin's sum over
{\em all} possible self-adjoint extensions.   Taking eq. (\ref{thet})
into account one can then compute  $\langle \tilde{S}_m\rangle$  in eq.
(\ref{smc}) exactly as before and obtain again our result (\ref{sm}) by
introducing now the variable $z = e^{i\theta'_m}$.  Thus, the
``pragmatic'' approach of the authors of Ref.\cite{ASV_00} yields
exactly the same average partial wave S-matrix as does our approach
based upon a straightforward application of mathematically rigorous but
perhaps less familiar techniques. Indeed, from (\ref{kap}), one can see that choosing one
specific self-adjoint extension amounts to making the bound states
spectrum unique by  specifying the position of {\em one} level in the
spectrum \cite{PP}. We thus have demonstrated, with the elastic
scattering solutions of Ref. \cite{ASV_00}, that our averaging procedure
leading to absorption and ultimately to the correct classical limit 
can be viewed as an average over all ways of specifying a unique
quantum-mechanical bound state spectrum.


\begin{thebibliography}{99}
\bibitem{singular}
By definition, a cylindrically symmetrical singular potential $V(r)$ is
characterized by the property \[\stackrel{\textstyle {\rm
lim}}{r\rightarrow 0}[r^2V(r)]\neq 0\,.\]  For an attractive singular
potential the possibility, classically, that the particle ``fall" into
the potential hole~\cite{DS_97} corresponds, in the quantized theory,
to an ambiguity in the choice of the radial wave function, which
implies the impossibility to obtain unambiguously the scattering phase
shift.  The mathematical characterization of this ambiguity lies in the
statement that ``Singular differential operators are either
self-adjoint or give rise to a well-defined class of self-adjoint
extensions  in the Hilbert space of square-integrable functions
$L_2(0,\infty)$." taken from Ref.~\cite{Meetz}.
\end{thebibliography}
\end{document}